\documentclass[usenatbib]{basi}
\usepackage[british]{babel}
\begin{document}
\title[BASI sample paper]{Comparative analysis of modern empirical spectrophotometric atlases with multicolor photometric catalogues\thanks{By
            Elena Kilpio, email: \texttt{lena@inasan.ru}}}
\author[E.~Yu.~Kilpio et~al.]%
       {E.~Yu.~Kilpio$^1$\thanks{email: \texttt{lena@inasan.ru}}, O.~Yu.~Malkov$^{1,2}$,
       and A.~V.~Mironov$^{3}$\\
       $^1$Institute of Astronomy, Russian Acad. Sci., 48 Pyatnitskaya, Moscow 119017, Russia\\
       $^2$Faculty of Physics, Moscow State University, Leninskie Gory, Moscow 119991 GSP-1, Russia\\
       $^3$Sternberg Astronomical Institute, Moscow State Univ.,13 Universitetskij Prosp., Moscow 119992, Russia}

\pubyear{2012}
\volume{00}
\pagerange{\pageref{firstpage}--\pageref{lastpage}}

\date{Received \today}

\maketitle
\label{firstpage}

\begin{abstract}

We present the results of the comparative analysis of the most
known semi-empirical and empirical spectral atlases that was carried
out using the data from the WBVR photometric catalogue.
The results show that standard error of synthesized stellar magnitudes
calculated with SEDs from best spectral atlases reaches 0.02 mag.
It has been also found out that some of modern spectral atlases are
burdened with significant systematic errors. The agreement for
the 5000-10000 A spectral range is rather satisfactory, while there
are problems for wavelengths shorter than 4400 A.

\end{abstract}


\section{Introduction}\label{s:intro}

For a great variety of astrophysical applications it is extremely
important to know (at least relative) spectral energy distribution
(SED) for
\begin{itemize}
  \item as many stars as possible, and
  \item as many types of stars as
possible.
\end{itemize}
 Semi-empirical spectrophotometric atlases are designed to
meet the latter requirement, meanwhile to satisfy the former one a
large number of empirical atlases are constructed.

The Asiago Database of Spectroscopic Databases (Sordo \& Munari,
2006) includes about 300 empirical stellar spectral atlases,
published from the late sixties to 2005. The majority of them covers
a very restricted spectral range or/and contains too small number of
stars. Less than a dozen of atlases contain large enough number of
objects (at least hundreds) and provide SEDs for spectral range of at
least 3000-9000 A. Among them are Pulkovo Spectrophotometric
Catalogue, Alma-Ata atlas, NGSL, ISO-SWS atlas and some others.
However, a comparison of data from these atlases for common stars
shows numerous and significant disagreements, especially for UV
spectral range.

On the other hand, multicolor high-precision photometry for much
larger number of stars is available in different photometric
systems. The most precise photometry can be found in
Hipparcos/Tycho-2 catalogues and in ground-based WBVR catalogue of
northern stars. So, the reliability of data in stellar spectral
atlases can be checked by a comparison of magnitudes, calculated
with methods of synthetic photometry, with catalogued data.

Here we present the results obtained using the observational
data from the Catalogue of WBVR magnitudes of Bright Northern Stars (Kornilov et al. 1991).
This catalogue provides the four-colour WBVR photoelectric magnitudes
for 13586 northern sky objects ($\delta>-15$) obtained at
the high-altitude observatory in Kazakhstan using a four-channel
photometer attached to the 0.5m reflector. The mean error in V ($\sigma$V) for nonvariable stars in the catalogue is $0.003^m$.
Limiting stellar magnitude is $V=7.2^m$. The catalogue is complete up to
$V=7^m$. The WBVR catalogue can be considered as a photometric
catalogue of Hipparcos level accuracy.

\section{Semi-empirical atlases}

\begin{figure}
\centerline{\includegraphics[width=9cm]{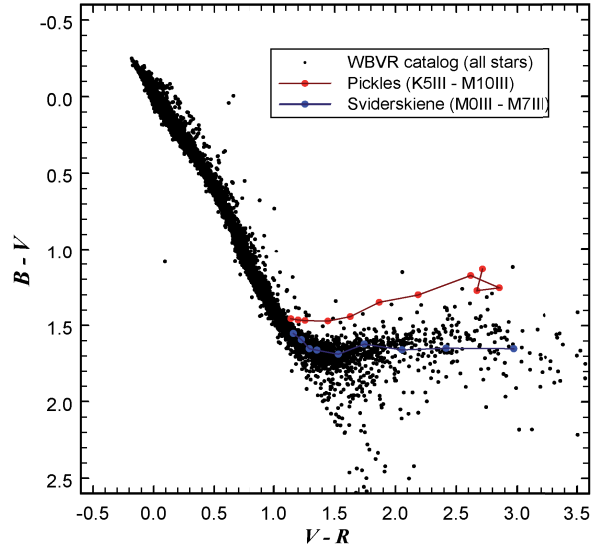}}
\caption{Synthetic B,V,R photometry, calculated from Pickles (1998)
and Sviderskiene (1988) data for red giants, together with
observational data from WBVR catalogue (Kornilov et al. 1991).
\label{f:fig1}}
\end{figure}

Only two of semi-empirical atlases among those contained in The Asiago Database
cover rather wide spectral range: Sviderskiene (1988) (1200-10500 A)
and Pickles 1998 (1150-25000 A). The semi-empirical stellar spectral
atlas of Pickles (1998) is widely used and remains one of the best
spectral atlases, however, as our analysis shows, synthetic
magnitudes calculated with Pickles data appreciably differ from
observed ones for some types of stars. Particularly (see Fig.\ref{f:fig1}),
B,V,R magnitudes of K4 and later giants predicted by Pickles (1998),
deviate significantly from ones listed in WBVR catalogue (Kornilov
et al. 1991) while the data from Sviderskiene (1988) fit them better.

\begin{figure}
\centerline{\includegraphics[width=8.5cm]{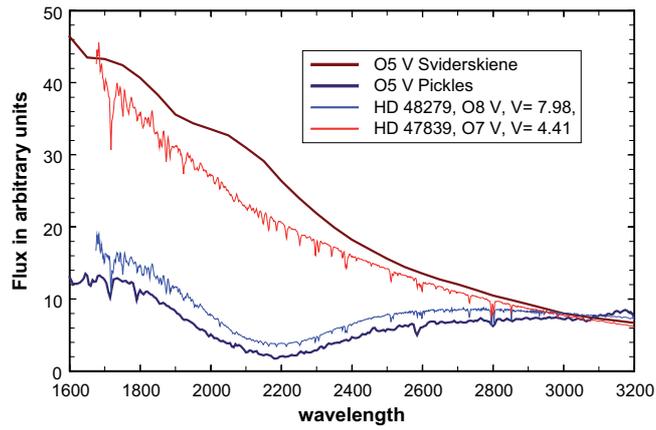}}
\caption{Data for O5V type star according to Pickles (1998)
and Sviderskiene (1988). Spectra of two real O type stars are also given.
\label{f:fig2}}
\end{figure}

It should be noted also (see Fig.\ref{f:fig2})
that data for O-type stars in the Pickles (1998) atlas
are not corrected for interstellar extinction carefully or specifically
enough. Pickles curve reproduces a SED for the faint, presumable reddened
HD~48279, while Sviderskiene (1988) data show a good agreement
with the brighter, presuimably unreddened HD~47839.

\section{Empirical spectral atlases}

Early empirical atlases (e.g.,
Breger (1976), Gunn and Stryker (1983)) contained few hundreds of stars.
In late 80th more representative atlases came up to take their
place: Spectrophotometric Catalogue of Stars (hereafter SCS) by
Kharitonov et al. (1988), Sternberg Spectrophotometric Catalog by
Glushneva et al. (1982-1984)
with its IR extension Moscow Spectrophotometric Catalog by Glushneva
et al. (1980-1991). The most precise data on 238 secondary
spectrophotometric standards (hereafter SSS) were collected by
Glushneva et al. (1992). Pulkovo Spectrophotometric Catalog
published by Alekseeva et al. (1996, 1997) should also be mentioned
among the most representative and precise spectrophotometric catalogues.

\begin{figure}
\centerline{\includegraphics[width=9cm]{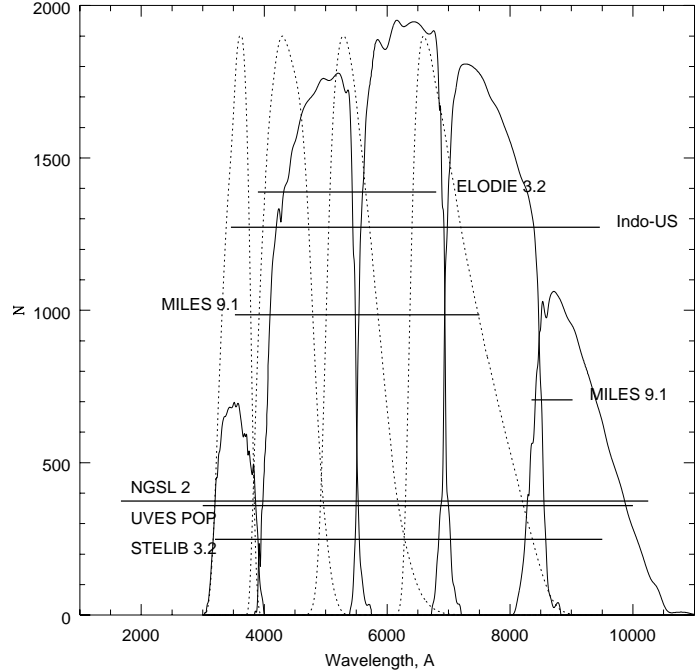}}
\caption{Horizontal lines represent modern empirical
spectrophotometric atlases: number of stars vs. spectral coverage.
SDSS (solid) and WBVR (dashed) response curves are also shown.
\label{f:fig3}}
\end{figure}

\begin{table}
\caption{Empirical spectral atlases}
\begin{tabular}{lrcl}
\hline
Name            & No of stars & Spectral      & Reference       \\
                &             & range         &                 \\
\hline
ELODIE 3.2      & 1388        & 3900-6800     & Wu et al. (2011)               \\
\multicolumn{4}{r}{http://www.obs.u-bordeaux1.fr/m2a/soubiran/elodie\underline{ }library.html}\\
Indo-US / CFLIB & 1273$^1$    & 3460-9464     & Valdes et al. (2004)           \\
\multicolumn{4}{r}{http://www.noao.edu/cflib/}\\
MILES 9.1       &  985        & 3525-7500     & Falcon-Barroso et al. (2011) \\
                &  706        & 8350-9020     &                            \\
\multicolumn{4}{r}{http://miles.iac.es}\\
NGSL 2          &  374        & 1670-10250    & Heap and Lindler (2007)        \\
\multicolumn{4}{r}{http://archive.stsci.edu/prepds/stisngsl/}\\
STELIB 3.2      &  249        & 3200-9500     & Le Borgne et al. (2003)        \\
\multicolumn{4}{r}{http://webast.ast.obs-mip.fr/stelib}\\
UVES POP        &  359$^2$    & 3000-10000    & Bagnulo et al. (2003)            \\
\multicolumn{4}{r}{http://www.sc.eso.org/santiago/uvespop/}\\
\hline
SCS             &  1147       & 3225-7575     & Kharitonov et al. (1988, 2011) \\
\multicolumn{4}{r}{VizieR: III/202}\\
SSS             &   238       & 3200-7600     & Glushneva et al. (1992)        \\
                &    99$^3$   & 6000-10800    &                              \\
\multicolumn{4}{r}{VizieR: J/A+AS/92/1}\\
Pulkovo         &   679       & 3200-7350     & Alekseeva et al. (1996, 1997)  \\
                &   278$^3$   & 3200-10800    &                              \\
\multicolumn{4}{r}{VizieR: III/201}\\
\hline
\end{tabular}

$^1$ 885 stars cover the indicated spectral range, SEDs of other
stars have gaps.

$^2$ SEDs of some stars have gaps.

$^3$ ...of the preceding number of stars. \label{tab:atlases}
\end{table}


From the  Asiago Database (Sordo \& Munari 2006) we selected those atlases of
observed stellar spectra that contained enough stars and provided the data in wide spectral range.
These atlases are listed in Table~\ref{tab:atlases} (together with some earlier atlases) and shown in Fig.\ref{f:fig3}

\begin{figure}
\centerline{\includegraphics[width=9cm]{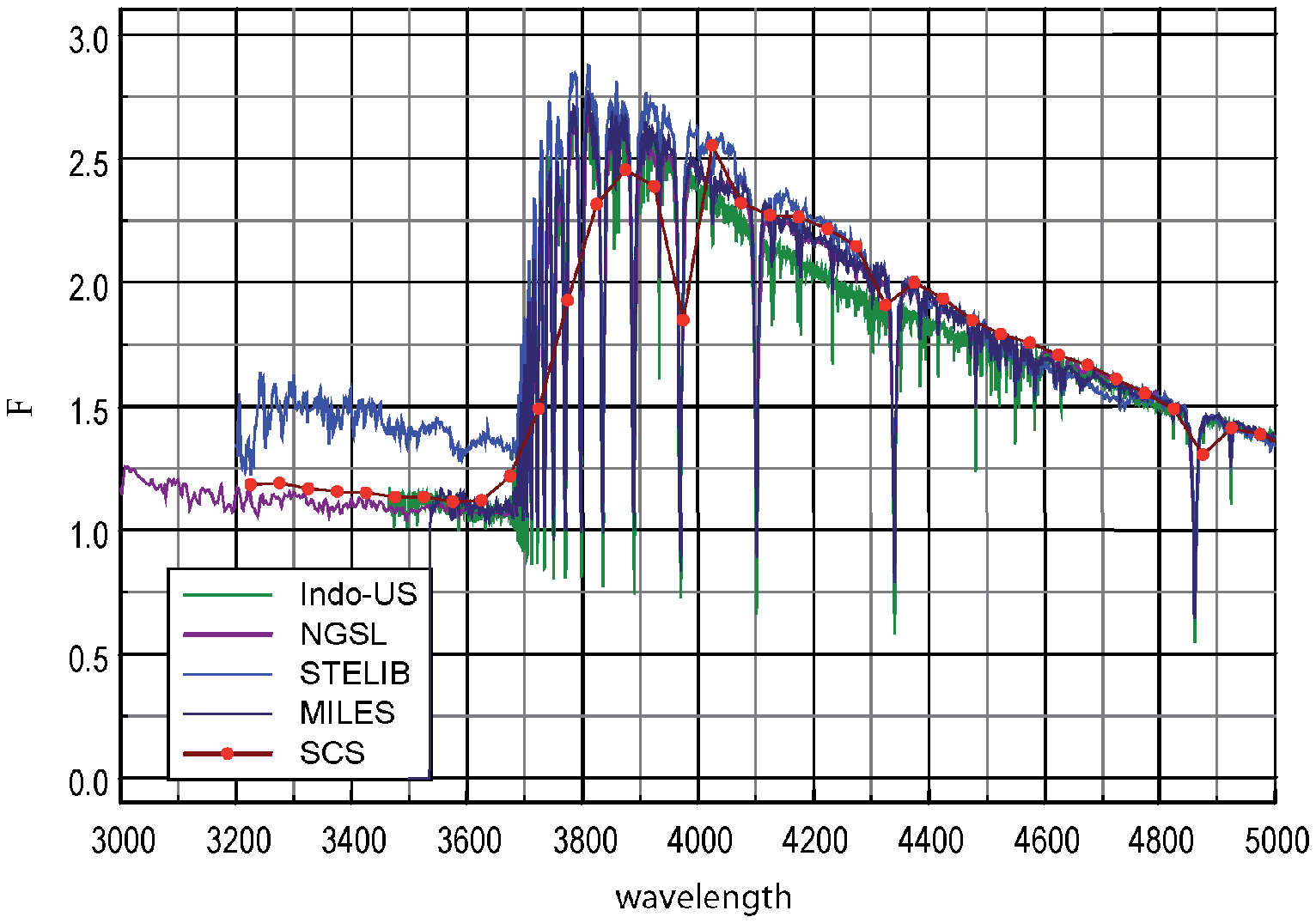}
            }
\centerline{\includegraphics[width=9cm]{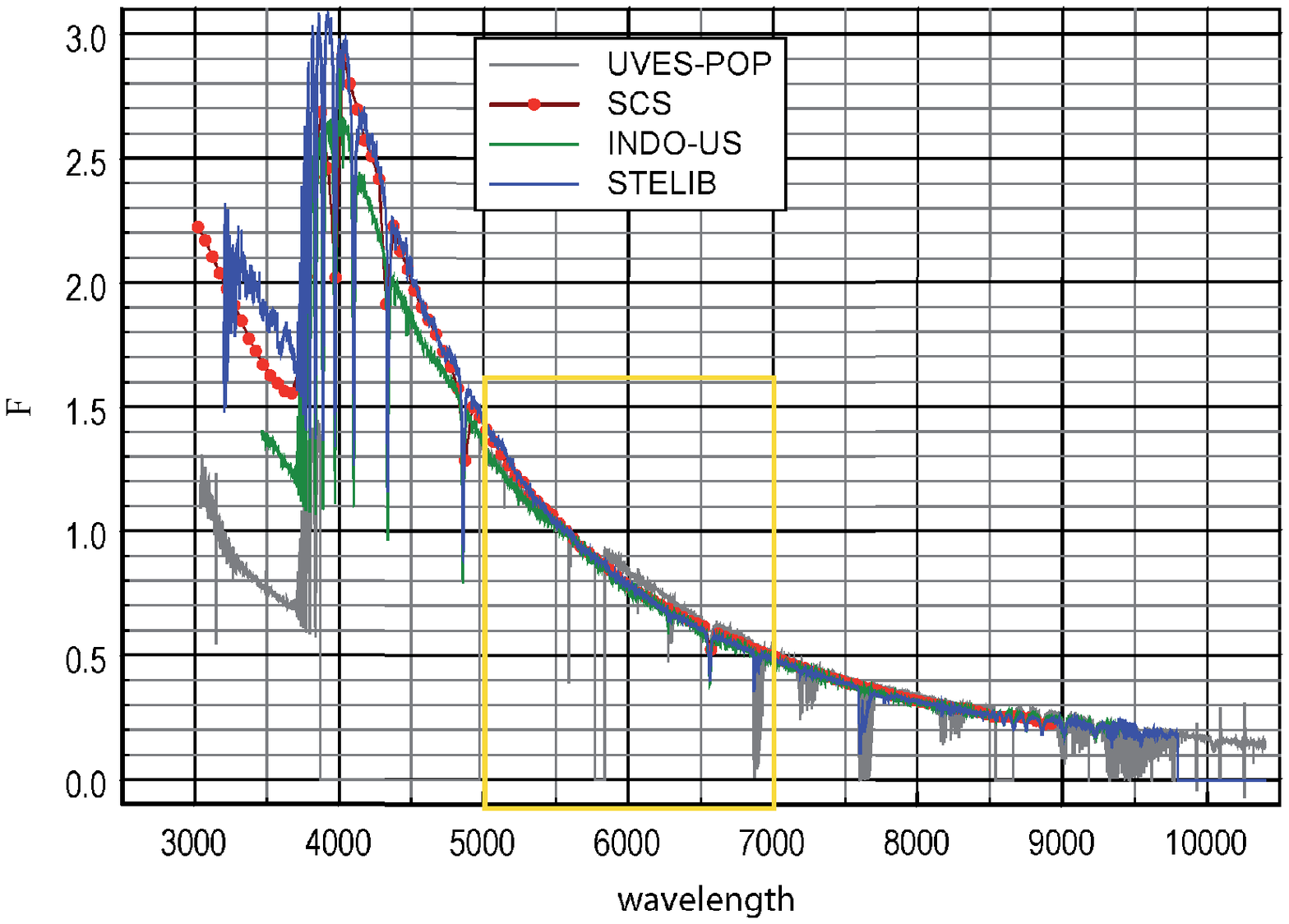}
            }
\caption{Spectral energy distribution of of HD~87737 (upper panel)
and $\alpha$ Leo (lower panel), presented in various atlases. Curves
are normalized at 5500 A. Both stars are located at high galactic
latitudes and both are not distant ones so the differences between
atlases we see here can not be explained as the effect of
interstellar extinction.\label{f:fig4}}
\end{figure}

\begin{figure}
\centerline{\includegraphics[width=9.5cm]{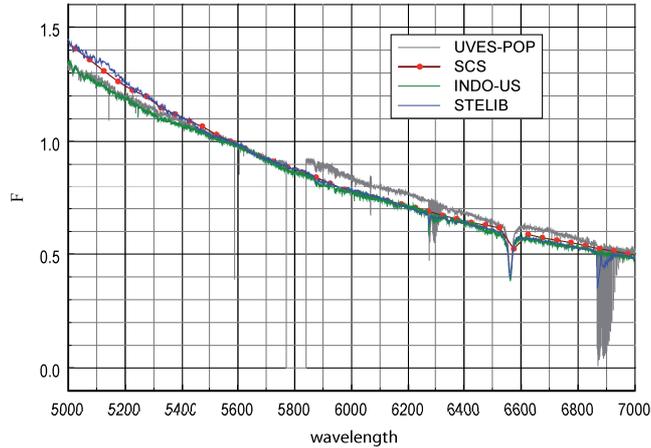}}
\caption{$\alpha$ Leo: the detailed view of the area marked in Fig\ref{f:fig4}
by the yellow rectangle is shown in details to illustrate the
example of bad sewing between two spectral ranges in UVES-POP.
\label{f:fig5}}
\end{figure}

Figures \ref{f:fig4}-\ref{f:fig5} illustrate rather typical results of comparison of atlases.
It should be noted that data presented in various atlases often
significantly differ even for bright stars, especially in the UV
range. Here one more effect can be also seen - sometimes there is bad
sewing between wavelength ranges in UVES-POP (see Fig\ref{f:fig4}).

Synthetic magnitudes and colors can be calculated for stars,
presented in empirical atlases, combining their spectral energy
distributions with response curves. Such magnitudes and colors in
WBVR system were calculated for stars, catalogued in (Kornilov et
al. 1991) and included in the atlases. Results of comparison for
various spectrophotometric atlases are presented in
Table~\ref{tab:wbv}.

\begin{table}
\caption{Comparison of synthetic and catalogued (Kornilov et al.
1991) colors}\label{tab:table1}
  \medskip
\begin{tabular}{l|rrrrrr}
\hline
              & UVES-POP & STELIB & MILES & Indo-US & NGSL & SSS      \\
\hline
$\sigma_{W-B}$&  0.043  &0.118  &0.192  &0.210  &0.028  &0.072   \\
No of stars   &  7  &13 &34 &76 &15 &231     \\
\hline
$\sigma_{B-V}$&  0.015  &0.093  &0.113  &0.087  &0.018  &0.020   \\
No of stars   &  6  &11 &34 &76 &15 &231     \\
\hline
\end{tabular}
\label{tab:wbv}
\end{table}

Preliminary analysis of modern empirical spectral atlases shows the
following.
\begin{itemize}
\item Accuracy of the ground-based SSS and
of the space-born NGSL atlases is comparable. Standard error of
synthetic photometry, calculated from the two SEDs, reaches
$0.^m02$.
\item The UVES-POP atlas is precise enough, however, sometimes
different spectral ranges are not sewed accurately.
\item Systematic errors of the other atlases are significant
(more than $0.^m1$).
\item Spectrum calibration problems in the UV spectral range remain
unsolved.
\end{itemize}

\section{Conclusions}

The comparative analysis of the most popular semi-empirical and
empirical spectral atlases has been carried out using the data from
WBVR photometric catalogue. Basing on the results of this analysis
we can make the following conclusions:

\begin{itemize}
  \item Standard error of synthesized stellar magnitudes, calculated with
SEDs from best spectral atlases, reaches 0.02 mag.
  \item Some of modern spectral atlases are burdened with significant systematic errors.
  \item SEDs from majority of atlases show satisfactory agreement for the 5000-10000 A
spectral range, but problems for wavelength shorter than 4400 A
remain.
\end{itemize}


\section*{Acknowledgements}

This work has been supported by Russian Foundation for Basic Research grants
09-02-00520, 10-07-00342 and 10-02-00426, by the Federal Science and Innovations Agency
under contract 02.740.11.0247, by the Federal target-oriented program ``Scientific and
pedagogical staff for innovation Russia'' (contract No. P1195),
and by the Presidium RAS program ``Leading Scientific Schools Support'' 4354.2008.2.
This research has made use of NASA's Astrophysics Data System Bibliographic Services (ADS).



\begin{thebibliography}{}

\bibitem[Alekseeva et al. (1996-1997)]{1997yCat.3201....0A}Alekseeva G. A., Arkharov A. A., Galkin V. D. et al.
1996, Baltic Astronomy 5, 603; 1997, Baltic Astronomy 6, 481

\bibitem[Bagnulo et al. (2003)]{2003Msngr.114...10B} Bagnulo S., Jehin E., Ledoux C. et al.
2003, Messenger, 114, 10

\bibitem[Breger (1976)]{1976ApJS...32....7B}
Breger M. 1976, ApJS, 32, 7

\bibitem [Falcon-Barroso et al. (2011)]{2011A&A...532A..95F} Falcon-Barroso J., Sanchez-Blazquez P., Vazdekis A. et al.
Astron. Astrophys. 532, id.A95

\bibitem[Glushneva et al.(1980-1991]{1980AZh....57.1003V}
Glushneva I.N., Doroshenko V.T., Fetisova T.S. et al.
1980, Soobsh.Gos. Astron. Inst. Shternberga 219,3;1980, Astron. Zh.
57,1003;1982, Trudy Gos.Astron.Inst. Shternberga 52, 182; 1983,
Trudy Gos.Astron.Inst.Shternberga 55, 84;1989, Trudy
Gos.Astron.Inst.Shternberga 61, 272;1991, Trudy Gos.Astron.Inst.
Shternberga 62,119

\bibitem [Glushneva et al. (1983)]{1983TrSht..53...50G}
Glushneva I.N., Doroshenko V.T., Fetisova T.S. et al.
1982, Spectrophotometry of Bright Stars (ed. Glushneva I. N.)
Moscow, Nauka, pp.3-252; 1983, Trudy Gos. Astron. Inst. Shternberga
53, 50; 1984, Trudy Gos. Astron. Inst. Shternberga 54, 3

\bibitem[Glushneva et al. (1992)]{1992A&AS...92....1G}
Glushneva I. N., Kharitonov A. V., Kniazeva L. N, Shenavrin,
V. I. 1992, Astron. Astrophys. Suppl. Ser., 92,~1

\bibitem[Gunn and Stryker (1983)]{1983ApJS...52..121G}
Gunn J. E., Stryker L. L. 1983, ApJS 52, 121

\bibitem [Heap and Lindler (2007)]{2007ASPC..374..409H}
Heap S. R., Lindler, D. J. 2007, ASPC 374, 409

\bibitem[Kharitonov et al (1988)]{1988scs..book.....K}
Kharitonov A. V., Tereshchenko V. M. and Knyazeva L. N. 1988,
Alma-Ata, Nauka, p. 484;
recent 3rd edition: 2011, Spectrophotometric Catalogue of Stars (ed.
Tereshchenko V. M.) AlmaAty, Kazakh. Univ., pp.4-304

\bibitem[Kornilov et al. (1991)]{1991TrSht..63....1K}
Kornilov V.G., Volkov I.M. et al. Catalogue of WBVR magnitudes of
Bright Northern Stars, Ed.-in-Chief Kornilov V.G., M.: Moscow State
Univ. Publ., 1991

\bibitem [Le Borgne et al. (2003)]{2003A&A...402..433L}
Le Borgne J.-F., Bruzual G., Pello R. et al.
2003, A\&A 402, 433


\bibitem[Pickles A.J. (1998)] {1998PASP..110..863P}
Pickles A.J. 1998, Publ. Astron. Soc. Pac. 110, 863

\bibitem[Sordo et al.(2006)]{2006A&A...452..735S}
Sordo R., Munari U. 2006, A\&A 452, 735

\bibitem[Sviderskiene Z.(1988)]{1988VilOB..80....3S}
Sviderskiene Z. 1988, Vilnius Obs. Bull. 80, 3

\bibitem [Valdes et al. (2004)]{2004ApJS..152..251V}
Valdes F., Gupta R., Rose J. A. et al.
2004, Astrop. J. Suppl. 152, 251

\bibitem[Wu et al. (2011)]{2011A&A...525A..71W} Wu Y., Singh H. P., Prugniel P. et al.
2011, A\&A, 525, A71

\end{thebibliography}
\end{document}